\documentstyle[twocolumn,eqsecnum,aps,epsf]{revtex}

\begin{document}
\draft
\title{Low-dimensional dynamics embedded in a plane  Poiseuille flow
turbulence
\\Traveling-wave solution is a saddle point?
}
\author{Sadayoshi Toh\cite{ma1} and Tomoaki Itano\cite{ma2}}
\address{
Division of Physics and Astronomy, Graduate School of Science \\
Kyoto University, Kyoto 606-8502, Japan\\
{\rm (May\ \ 1, 1999)}\\
\ \\
\begin{minipage}{14cm}
\rm \ \ \ \ The instability of a streak and its nonlinear evolution are investigated by direct
 numerical simulation (DNS) for plane Poiseuille flow at $Re=3000$. 
It is suggested that there exists a traveling-wave solution (TWS). The TWS
 is localized around  one of the two walls
and notably resemble to the coherent structures observed in 
experiments and DNS so far.    
 The phase space structure around this TWS is 
similar to a saddle point. 
Since the stable manifold of this TWS is extended close to the quasi two 
 dimensional (Q2D) energy axis,  the approaching process toward 
 the TWS along the 
stable manifold is approximately described as the instability of 
the streak (Q2D flow) and the succeeding nonlinear evolution.  
Bursting  corresponds to the escape from the TWS along the unstable manifold. 
These manifolds  constitute part of basin boundary of the 
turbulent state. 
\end{minipage}
}
\maketitle
\narrowtext
In wall-turbulent shear flows, turbulence production is believed 
to occur or at least to be triggered in the near-wall region in a 
deterministic and intermittent way, and also to be related to coherent 
structures: bursting\cite{General}. 
Although the mechanism of the bursting has not been well
understood, the coherence of this process suggests that 
a low-dimensional dynamics is embedded even in fully-developed 
turbulence\cite{modelL}. 
This low-dimensionality observed in subcritical flows\cite{Sub}
 is quite different from weakly  nonlinear stage of supercritical flows
such as B\'enard convection\cite{Super}, because the fully-developed 
turbulence coexists. 
That is, the low-dimensional dynamics should be connected to the 
huge-dimensional dynamics globally. 
\par 
This global structure seems to be hopelessly complicated. 
 However, a minimal flow unit  for maintaining turbulence 
was found by  Jin\'emez and Moin \cite{Minimal} in plane Poiseuille flow. 
 Owing to this work we can focus on the elementary process of turbulence
 in the near-wall region from the  perspective of dynamical systems
 without being bothered by spatial coexistence of  various stages. 
 Hamilton et al.\cite{SSP} examined the minimal flow unit
 of plane Couette flow
 and found that turbulence is sustained by not random but  quasi-cyclic 
 process
 passing quiescent and activated periods in turn. The quiescent
 period is relatively laminar where coherent structures, the so-called
 streaks, are formed, develop and become unstable.  Then their
 instability triggers an activated  or turbulent period.
 They called this process self-sustaining process (SSP).   
 Note that the Reynolds numbers ($Re$) used in their work are quite low, 
 so they observed spatial chaos rather than  fully-developed turbulence.
\par
 This picture seems to work well even in realistic situations.
 There are, however, still open questions.
 One is on the nonlinear evolution after the instability of streaks and
 another is  why the unstable streaks are regenerated. 
 The low-dimensional model introduced heuristically by Waleffe\cite{W1} 
 has partially answered to these.
 The keys are the existences of steady solutions and a homoclinic orbit.
 This model, however, is wholly low-dimensional and  not related to 
fully-developed turbulence.
\par
  Recently, Waleffe obtained  traveling-wave solutions (TWS)
  in plan Poiseuille flow  by continuing from  steady solutions in 
  plane Couette flow for stress boundary conditions\cite{W2}.  
  These solutions are quite similar in shape to 
  the coherent structures educed by Joeng et al.\cite{Hussain1} 
  in  direct numerical simulation (DNS), although the boundary
  conditions are different. 
\par
  DNS and experiments have revealed the existence of coherent
 structures in the near-wall  region that are intimately linked to 
  the maintenance  of turbulence.
 In fact the sinuous evolution of a streak observed in DNS appears to 
 trigger off the formation of longitudinal coherent structures and 
 the succeeding bursting. This process is also the main part of 
 the SSP picture. 
 Schoppa and Hussain\cite{SP0} showed  
 that  the instability of ejected low speed streaks directly generates 
 new streamwise vortices, internal shear layers and arch vortices 
 by examining stability of a vortex-less streak\cite{Streak}.
 The resulting 3D vortex geometry is
 identical to that of the dominant coherent structures educed from 
 the near-wall turbulence, too.
 They concluded that vortex-less streaks are the main breeding ground for
 new streamwise vortices, commonly accepted as dominant in turbulence 
 production. 
\par
 In this paper, we try to elucidate the relation of
 the conceptual ingredients of the  SSP picture such as the instability of 
 streaks and  coherent structures with TWS,  homoclinic nature, etc. 
 in a more realistic situation by means of DNS. 
\par 
The numerical scheme adopted here is basically the same as those of  
 Kim et al.\cite{Kim}. 
 For all simulations reported, we use periodic boundary conditions 
 in $x$ and $z$ with period lengths $L_x=\pi\sim 420^+$ and 
 $L_z=0.4\pi\sim 170^+$, and apply  no-slip boundary conditions 
 at $y=\pm 1$: $x$, $y$ and $z$ represent the streamwise, 
 normal to the walls and spanwise directions, respectively. 
 The value of Reynolds number is fixed to 3000. 
Though somewhat large, this system belongs to the minimal flow unit
 \cite{Minimal}.
 The flow is maintained by a constant volume flux.
As aliasing errors are removed by 1/2 phase-shift method, the effective 
modes are $30\times 30$ Fourier modes in $x$ and $z$, and 65 Chebychev 
polynomials in $y$. 
\par
In Fig.~\ref{ReStress}, the evolution of  Reynolds stress
integrated over the horizontal plane, 
$\vert\langle u_xu_y\rangle(y,t)\vert$ 
is shown. Turbulent 
fluctuations are produced close to the walls and ejected into
the center region. These sudden production and ejection of fluctuations 
are typical bursting processes\cite{Lumly2}. 
Bursting on both walls seems to occur independently. 
In the lower half region,  an especially 
quiescent period where the flow is laminarized and a prominent streak 
develops, lasts about ten times as long as ordinary one.
Even in this period, bursting continues to occur in the upper region.     
This quasi-cyclic and intermittent occurrence of  bursting is
reminiscent of the chaotic homoclinic orbit\cite{GH}. 
\par
To describe the low-dimensional dynamics suggested,
we introduce the 2D phase space spanned by the quasi-2D (Q2D) and 3D 
components of kinetic energy of the normal velocity, $u_y$, that is,  
 $E_y^{\mbox{\scriptsize Q2D}}=\int_V(u_y^{\mbox{\scriptsize Q2D}})^2
\mbox{d}V/2$  
and $E_y^{\mbox{\scriptsize 3D}}=\int_V (u_y^{\mbox{\scriptsize
3D}})^2\mbox{d}V/2$,
where $V$ denotes the whole volume of the computation. 
Here we decompose the velocity field into two parts: 
Q2D flow $\mbox{\boldmath $u$}^{\mbox{\scriptsize Q2D}}(y,z,t)$,
and 3D flow $\mbox{\boldmath $u$}^{\mbox{\scriptsize 3D}}(x,y,z,t)$. 
The latter is constructed by Fourier modes only with non-zero streamwise 
wave number. 
The velocity field restricted in Q2D space, is damped to the laminar 
plane Poiseuille flow monotonically for all  $Re$\cite{Q2Ddynamics}. 
In this sense, there is no fixed point on the Q2D axis except for the
origin.\par
Hamilton  et al.\cite{SSP}  defined the streak and the longitudinal
vortex as the $x$ component($u_x^{\mbox{\scriptsize Q2D} }$) of Q2D flow and its $y,z$ components 
$(u_y^{\mbox{\scriptsize Q2D} },u_z^{\mbox{\scriptsize Q2D} })$,
respectively for the minimal Couette turbulence. 
They also regarded  3D flow  as turbulent components.
The simplicity of these definitions is a little bit curious 
because coherent structures are observed in the near-wall region and
also three-dimensional. 
These definitions, however, conceptually work well.
This suggests that quiescent stages or generating processes of coherent 
structures are well described in a low dimensional phase space. 
\par
  To see the global behavior of the system in phase space, 
  we examine the nonlinear evolutions of small 3D disturbances superposed to 
  an artificially  obtained Q2D velocity field that seems to be a
  prototype of a well-developed streak, as done by Schoppa and 
 Hussain\cite{SP0}.
They used an analytic streak solution without longitudinal vortex\cite{Streak}.
In contrast, our streak solution  is constructed by removing 
the $x$-dependent modes, i.e. the 3D component, 
from the streak-dominated velocity field obtained in DNS at $t=337$ 
in Fig.~\ref{ReStress} in order to compare 
the obtained results with the real situation.
\par
The initial condition is  as follows: 
\begin{equation}
\mbox{\boldmath $u$}(\mbox{\boldmath $x$},0)
=\mbox{\boldmath $u$}^{\mbox{\scriptsize Q2D}}(y,z,0) + 
\sqrt{F_{ac}}\frac{||u_y^{Q2D}(0)||}{||u_y^{\mbox{\scriptsize 3D}}(0)||}
\mbox{\boldmath $u$}^{\mbox{\scriptsize 3D}}(\mbox{\boldmath $x$},0),
\end{equation} 
where  $||f||\equiv (\int_V f^2\mbox{d}V)^{1/2}$ and 
$\mbox{\boldmath $u$}^{\mbox{\scriptsize 3D}}(\mbox{\boldmath $x$},0)$
 is a solenoidal random vector field  with a given broad spectrum and 
random phase. 
 The following results are qualitatively independent of the form 
of $\mbox{\boldmath $u$}^{\mbox{\scriptsize 3D}}(\mbox{\boldmath $x$},0)$.
\par
We examine the dependence of the evolutions on the the relative
amplitude, $F_{ac}$ of the 3D, i.e., $x$-dependent disturbance.
In Fig.~\ref{TempEvolNORM},  the evolutions of
$E_y^{\mbox{\scriptsize Q2D}}(t)$ and 
$E_y^{\mbox{\scriptsize 3D}}(t)$
 for several values of $F_{ac}$ are shown.
After initial transient stage where the most growing mode is selected,
 3D disturbance ($E_y^{\mbox{\scriptsize 3D}}(t)$) seems to grow 
exponentially and then saturate to an equilibrium state in damped-oscillating. 
Finally the disturbance chooses either the route to laminar plane Poiseuille
flow or that to turbulent state depending on whether the value of $F_{ac}$ is 
smaller  or larger than the critical value. 
We refer to this critical value, which is close to $F^{I}_{ac}=9.116010224\times
10^{-8}$,  as $F_{ac}^c$. 
 The passing time till the final abrupt damping or growth 
gets longer as $F_{ac}$ approaches $F^c_{ac}$.
The exponential growth in the initial stage corresponds to the 
instability of the streak, while the streak 
($E_y^{\mbox{\scriptsize Q2D}}(t)$) continues to be damped. 
\par
Figure~\ref{PhaseSpace} shows the evolutions in the phase space spanned  by
$E_y^{\mbox{\scriptsize Q2D}}$ and $E_y^{\mbox{\scriptsize 3D}} $ for
several values of $F_{ac}$. 
It is easy to see that a fixed point like a saddle-focus exists. 
The trajectory is  oscillating but approaches 
monotonically  the expected fixed point.
\par
The difference between 
two solutions for different $F_{ac}$ increases exponentially with 
 the roughly constant growth rate, $\sigma_+=0.037$  even in  
the approaching period.
This suggests that the evolution of $E_y^{\mbox{\scriptsize 3D}}$ 
is described as a motion around a saddle point like
$(dX/dt,d(Z_r+iZ_i)/dt)=(\sigma_+ X,(-\sigma_-+i\omega)(Z_r+iZ_i)))$
,where $i$ is imaginary unit and $\sigma_+, \sigma_-$ are positive constants.
\par
The complex damping rate could be also estimated, but the existence of 
another monotonically-damping mode makes an estimate difficult
(see Fig. \ref{TempEvolNORM}). A rough estimate shows that the ratio
of these two damping rates is about 2: the damping rates are  -0.004 for
the real mode and -0.008 for the complex one. 
Therefore the  dimension of the contracted space around the fixed point 
is at least larger than 4. 
This dynamics is  more complicated than  Waleffe's model.
\par
Both the stable and unstable manifolds of the saddle point
constitute a separatrix that separates  turbulent  and laminar 
states. The stable manifold is extended close to the Q2D axis. 
Because of this closeness, the stability of streaks, i.e., of a Q2D
solution, is well recognized even in fully-developed wall turbulence 
like in SSP.
\par
We infer that for $F^c_{ac}$ the trajectory reaches the fixed point.
In physical space, the solution corresponding to this point 
has a 3D shape and  moves steadily with the velocity $v=0.75 \pm 0.05$ 
in the streamwise direction. Thus the fixed point must be TWS.
In this sense, this solution is not a fixed-point but a periodic orbit. 
Since we cannot yield the TWS by DNS in a strict sense anyway,
 we regard the  saturated state obtained for
$F^{I}_{ac}$ as TWS hereafter.   
\par
  This TWS is notably resemble in shape to those obtained 
by Waleffe for stress boundary conditions as can be seen in
Fig.~\ref{Shape}, although  our $Re$
 is  about ten times as large as Waleffe's.
The streamwise wave length of Waleffe's solution is almost the same as ours
 and the spanwise periodicity  is  1.67 times as long as ours.
Note that 
while Waleffe's are symmetric with respect to the centerline of the channel, 
 our TWS is confined to the lower wall (see Fig.~\ref{Meanflow}).
 Our TWS is also so tall in height that it reaches to the upper boundary 
of the  log-law region  and is not localized to the near-wall region.      
Furthermore, in wall turbulence an interval between adjacent streaks 
is about $100$ in wall unit on average.
 If TWS is linked to the coherent structures, there could exist TWSs 
 with shorter height and/or spanwise periodicity than that of the TWS 
found by us. (See Fig.~\ref{Shape2}(c).  Two ejections are observed on
the upper wall.)  
This suggests that  other TWSs confined to the near-wall 
region  exist  and 
these TWSs should be common in realistic turbulence.
\par
 To compare the  quiescent period of the turbulence and 
the low dimensional dynamics around the TWS,
we also project the evolution of the turbulence 
onto the 2D phase
space as shown in  Fig.\ref{PhaseSpace}.
  Here we only use the energies contained 
in the lower half volume because the other turbulent evolution
occurs independently  in the upper half volume (see Fig.~\ref{Shape2}(c)). 
It is quite surprising that the projected trajectory is close to 
the unstable manifold, because wall turbulence is usually coherent or in 
order only in the near-wall region.
In the bursting period the system  goes away from the TWS along 
the unstable manifold. This suggests that the instability  not of a streak 
but of a TWS is the origin of the bursting process. 
In Fig.~\ref{Shape2} we show  the solutions on the
unstable manifold  and the trajectory of turbulence marked 
in Fig.~\ref{PhaseSpace}. From these figures, the resemblance is apparent. 
\par
The most unstable mode of the streak has the same wavelength as
 the TWS itself for $L_x=\pi$. This seems to imply that the instability of
streaks directly breed  coherent structures,  i.e., TWS. 
This is, however, not always true. Indeed, when $L_x$ is doubled, 
the most unstable mode to the streak is saturated to another TWS 
with double the periodicity: 
the streamwise wave 
numbers of the most unstable modes  for $L_x=\pi$ and $2\pi$ are the same. 
Moreover, we believe that  bursting is the escaping process 
from TWS along  the unstable manifold. Thus the stability of TWS 
is more important than that of streak. The further study of the former 
instability should elucidate   bursting, i.e.,  turbulence
generation and also the determination of streak intervals in wall turbulence. 
\par
As mentioned above, there may exist many TWSs. 
These TWSs may be connected with each others through turbulent or 
activated periods, although we have not understood how they are.
This generalized homoclinic or ``multi-clinic'' nature of wall
 turbulence seems to be the substance of coherent structures and
 quasi-cyclic evolution, SSP.
\par  
In the context of control of turbulence generation, the existence of
separatrix supports the laminarization of turbulent flow by
forcing the flow to be two-dimensional at least in the near-wall region
{\it e.g.} by means of riblets or suction\cite{Lumly2,SP1}.
\par
The computations have been performed on the NEC/SX4 of YITP, Kyoto Univ.

\newpage
\setcounter{figure}{0}

\vskip1ex
\centerline{ \epsfxsize=7.5cm \epsfysize=5cm \epsfbox{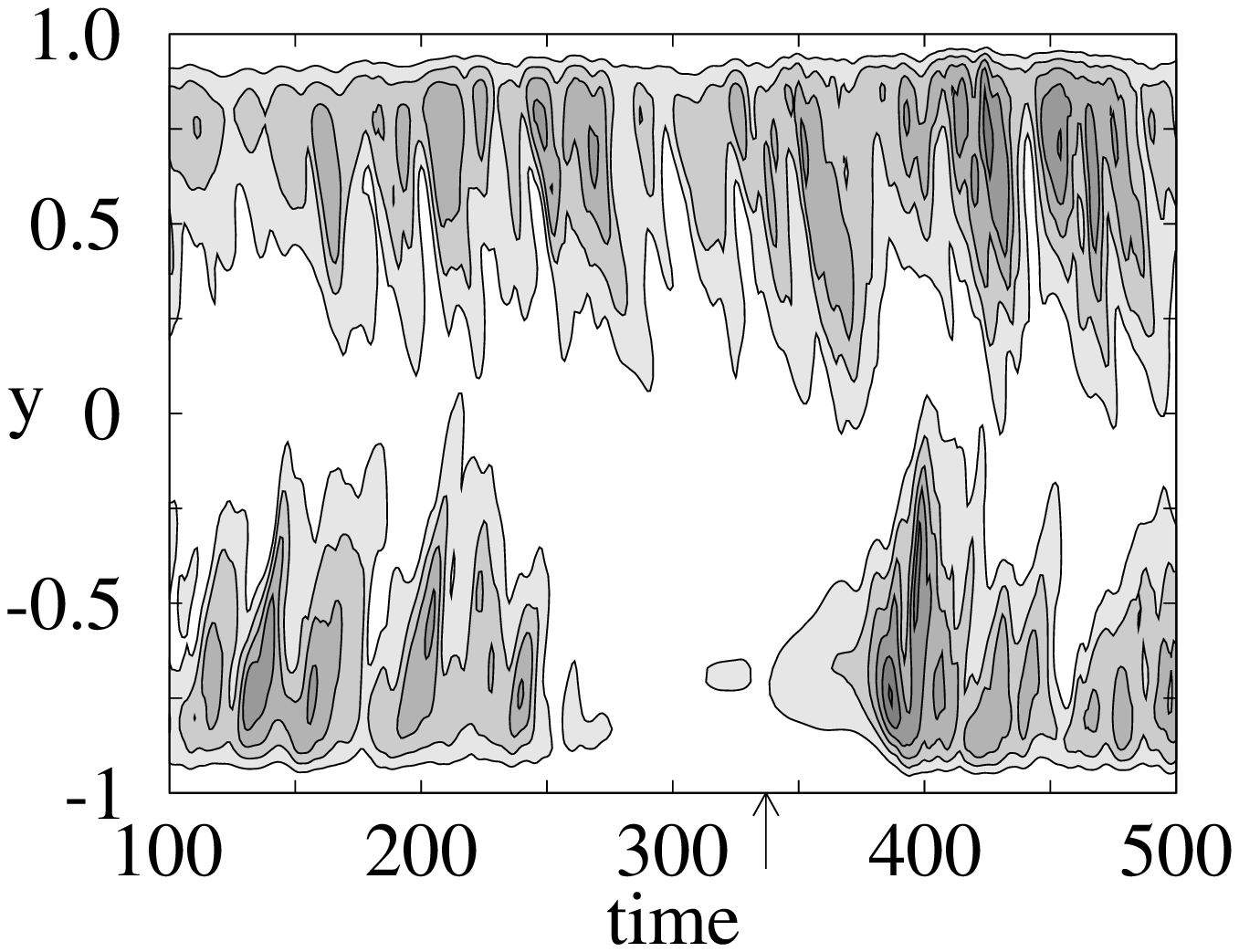} }
\begin{figure}
\caption{Evolution of Reynolds stress integrated over
 horizontal plane. Shade indicates  $\vert\langle u_xu_y\rangle(y,t)\vert\ge0.5$.}
\label{ReStress}
\end{figure}

\vskip1ex
\centerline{ \epsfxsize=7.5cm \epsfysize=5cm  \epsfbox{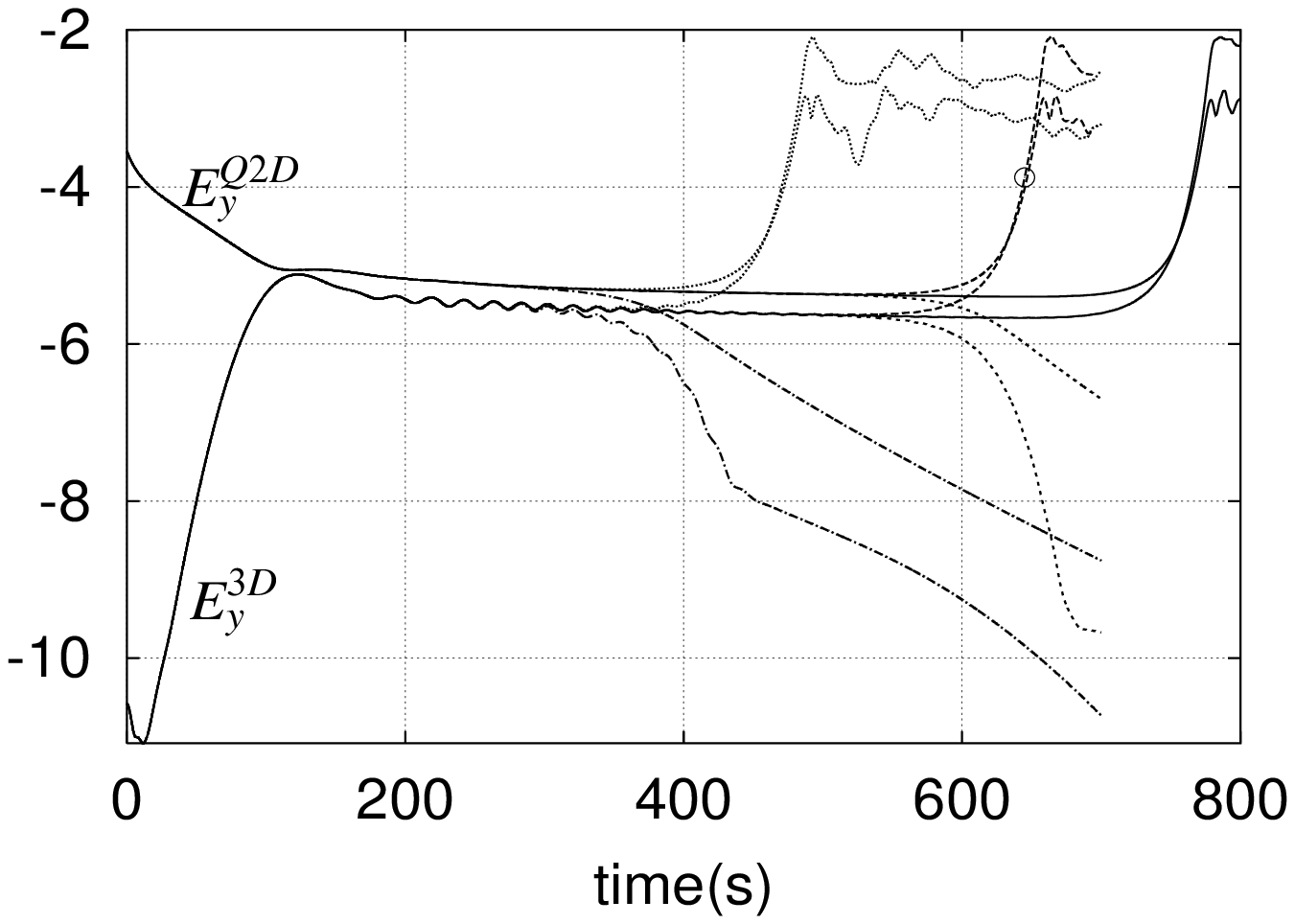} }
\begin{figure}
\caption{Linear-log plot of evolutions of $E_y^{Q2D} (t)$ and $E_y^{3D} (t)$ for
$F^{I}_{ac}=9.116010224\times 10^{-8}$ (solid line), 
$F^{II}_{ac}=9.1160104\times 10^{-8}$ (long-dashed line; 
$\circ$ indicates $t=645$), 
$F^{III}_{ac}=9.11601\times 10^{-8}$ (short-dashed line), 
$F^{IV}_{ac}=9.1162\times 10^{-8}$ (dotted line), 
$F^{V}_{ac}=9.114\times 10^{-8}$ (dash-dotted line).
}
\label{TempEvolNORM}
\end{figure}

\vskip1ex
\centerline{ \epsfxsize=8cm \epsfysize=5.5cm \epsfbox{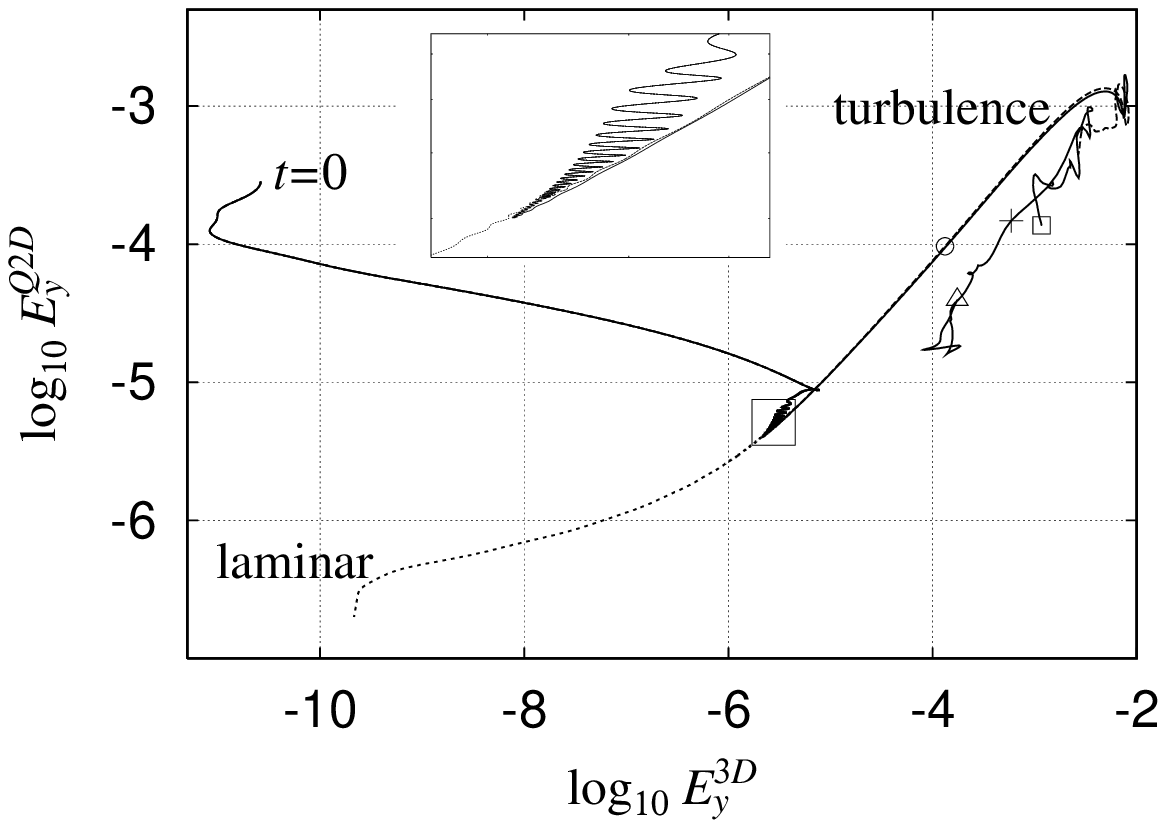} }
\begin{figure}
\caption{Log-log plots of evolutions of $(E_y^{Q2D} (t), E_y^{3D} (t))$ in 2D phase
 space. Correspondence between line and $F_{ac}$ is the same as that 
 in Fig.2.
 Thick solid line shows the evolution of the turbulence from $t=300
 (\triangle) $ to $t=450(\Box)$ via $t=380(+)$ in Fig.1.
 $\circ$ indicates $t=645$ for $F^{II}_{ac}$
.}
\label{PhaseSpace}
\end{figure}

\vskip1ex
\centerline{ \epsfxsize=8cm \epsfysize=6cm  \epsfbox{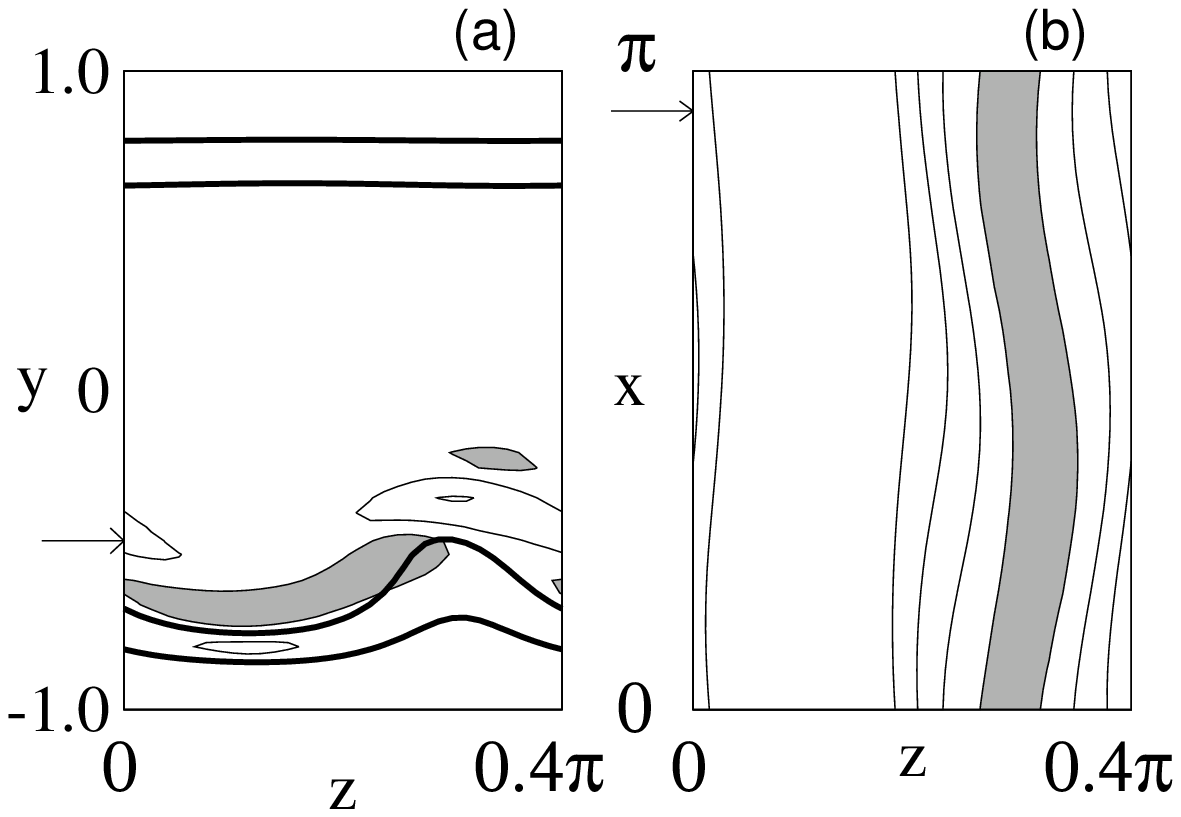} }
\begin{figure}
\caption{Snapshot of TWS at $t=600$ for $F^{I}_{ac}$.   
(a) longitudinal vorticity $\omega_x(y,z)$ at $z^+=398$.
(b) $u_x(x,z)$ at $y^+=71$.  Shade indicates  $u_x<0.65$ or $\omega_x<-0.05$.
Arrows indicate the positions of the section. 
Thick solid lines are for $u_x= 0.4$ and $0.6$.}
\label{Shape}
\end{figure}

\vskip1ex
\centerline{ \epsfxsize=7cm \epsfysize=5.5cm \epsfbox{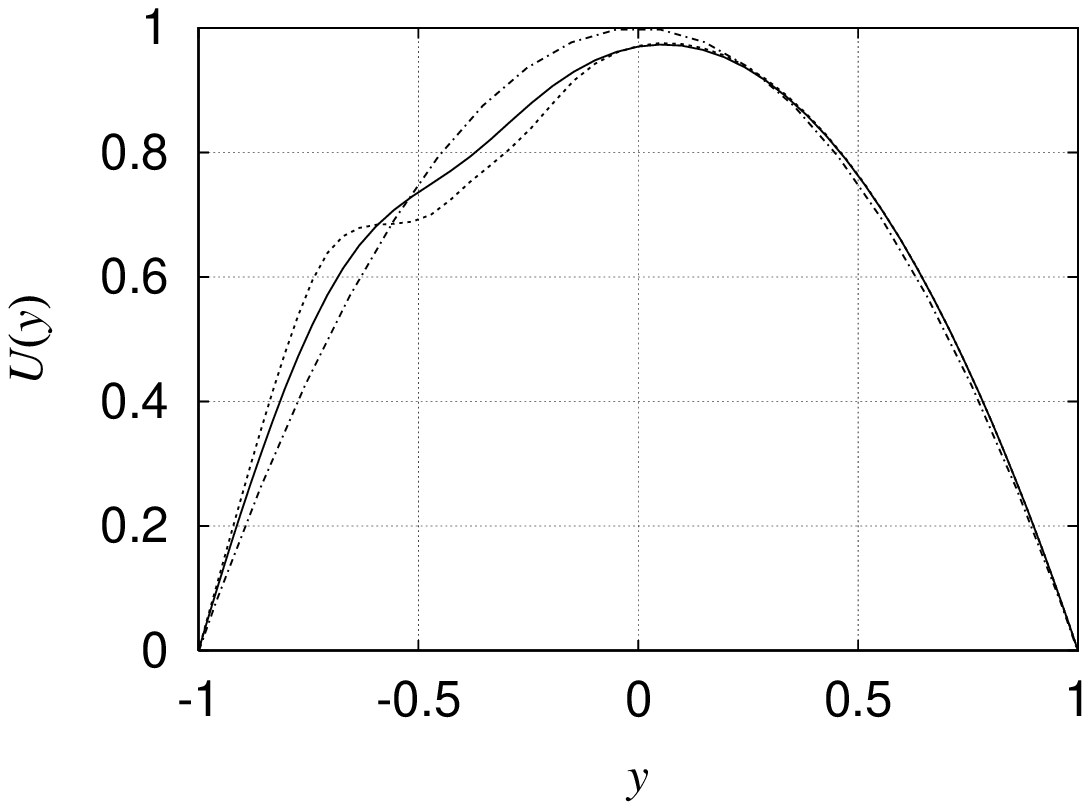} }
\begin{figure}
\caption{Mean streamwise velocity profile $U(y,t)$.
Solid line: TWS at $t=600$ for $F^{I}_{ac}$ in
Fig.2.
Dash-dotted line: laminar plane Poiseuille flow.
Dashed line: a snapshot close to the unstable manifold 
at $t=650$ for $F^{II}_{ac}$.}
\label{Meanflow}
\end{figure}

\vskip1ex
\centerline{ \epsfxsize=7.5cm \epsfysize=12cm  \epsfbox{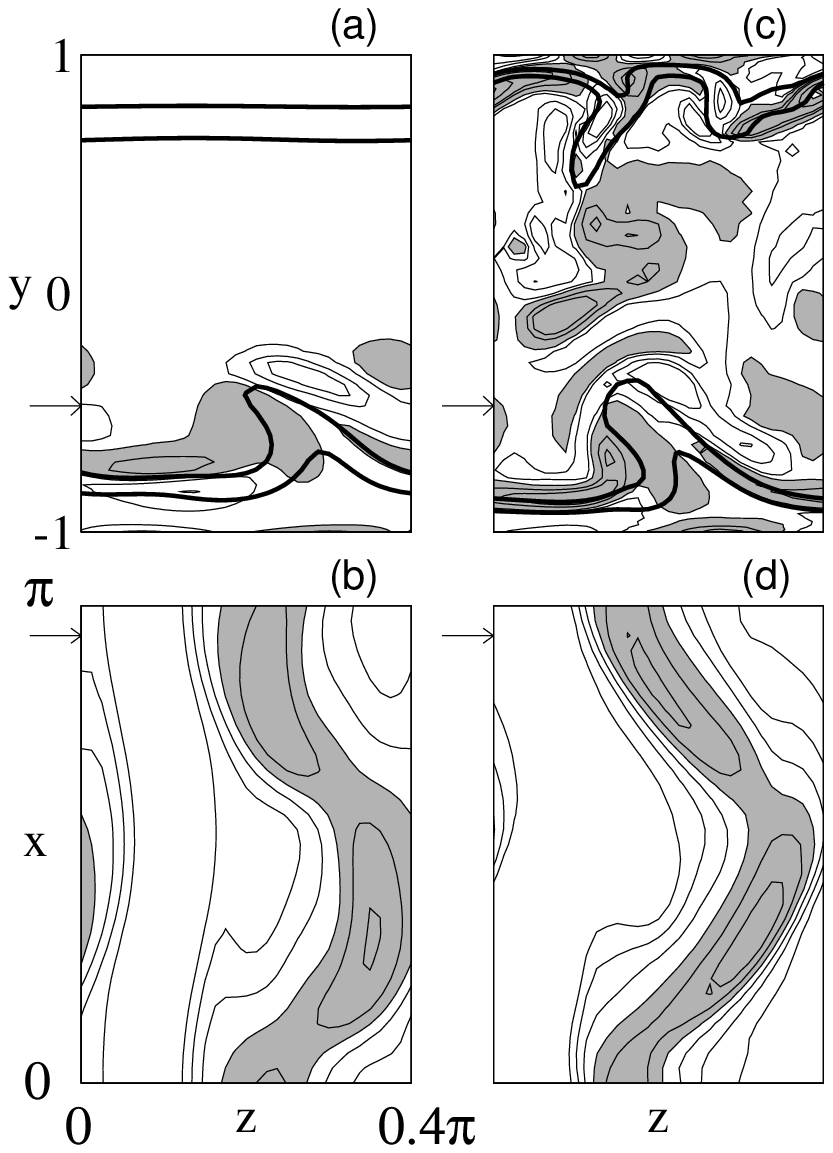} }
\begin{figure}
\caption{Snapshots. (a),(b): close to the unstable manifold 
($\circ$ in Fig.3), 
 (c),(d): in the bursting process of turbulence($+$ in Fig.3).
(a),(c) $\omega_x(x,y)$ at $z^+=398$. 
(b),(d) $u_x(x,z)$ at $y^+=71$.  Shade indicates  $u_x<0.65$ or
 $\omega_x <- 0.1$.  Thick solid lines are for $u_x= 0.4$ and $0.6$. 
 The phase of (c) and (d) is shifted for ease of comparison.}
\label{Shape2}
\end{figure}


\begin{references}
\bibitem[*]{ma1}  toh@kyoryu.scphys.kyoto-u.ac.jp
\bibitem[**]{ma2}  itano@kyoryu.scphys.kyoto-u.ac.jp
\bibitem{General} S. K. Robinson, Annu. Rev. Fluid Mech {\bf 23},601 (1991).
\bibitem{modelL}P. Holmes,J.L. Lumly, and G. Berkoozt, ``Turbulence,
 coherent structures, dynamical systems and symmetry''(1996)Cambridge
 Univ. Press.
\bibitem{Sub} O. Dauchot and P.  Manneville, J. Phys. II France {\bf
 7},371 (1997))
\bibitem{Super} M.C. Cross and P.C. Hohenburg, Rev. Mod. Phys. {\bf 65},
851 (1993). 
\bibitem{Minimal} J. Jim\'enez and P. Moin, J.\ Fluid  Mech.\ {\bf
 225},213 (1991).
\bibitem{SSP}J.M. Hamilton, J. Kim, and F. Waleffe, J.\ Fluid
 Mech.\ {\bf 287},317 (1995).
\bibitem{W1}F. Waleffe,  Phys.\ Fluids {\bf 7},883 (1995).
\bibitem{W2}F. Waleffe,  Phys.\ Rev.\ Lett.\ {\bf 81},1049 (1998).
\bibitem{Hussain1}J. Jeong, F. Hussain, W. Schoppa, and J. Kim,   J.\
 Fluid  Mech.\ {\bf 332},185 (1997).
\bibitem{SP0}
 W. Schoppa and F. Hussain, in Proceedings of the 29th AIAA Fluid
 Dynamics Confference, Albuqerque, NM, 1998
\bibitem{Streak} Their  streak solution is 
a steady solution only of the Euler equation but satisfies the 
no-slip boundary condition. 
\bibitem{Kim} J. Kim, P. Moin, and R. Moser,   J.\ Fluid  Mech.\ {\bf
 177},133 (1987).
\bibitem{Lumly2} J. Lumly and P. Blossey, Annu. Rev. Fluid Mech {\bf 30},311 (1998).
\bibitem{GH}J. Guckenheimer and P. Holmes, ``Nonlinear Oscillations,
 Dynamical Systems, and Bifurcations of Vector Fields'' (1983)Springer-Verlag.
\bibitem{Q2Ddynamics} In Q2D flow, $u_y$ and $u_z$ satisfy the 
2D Navier-Stokes equation without forcing.
\bibitem{SP1} W. Schoppa and F. Hussain,  Phys.\ Fluids {\bf 10},1049 (1998).
\end{references}
\end{document}